\documentclass[jap,
                    a4,twocolumn
                    ]{revtex4}

\usepackage{epsfig}
\usepackage{amsmath}
\usepackage{setspace}
\usepackage{subfigure}

\def\e{\begin{equation}}
\def\f{\end{equation}}

\begin{document}


\title{A microwave transmission-line network guiding electromagnetic fields through a dense array of metallic objects}

\author{Pekka~Alitalo, Sylvain~Ranvier, Joni~Vehmas, Sergei~Tretyakov}

\affiliation{Department of Radio Science and Engineering / SMARAD
Center of Excellence\\ TKK Helsinki University of
Technology \\P.O. Box 3000, FI-02015 TKK, Finland\\
}

\maketitle

\parskip 7pt

\vspace{0.5cm}
\begin{center}
\section*{Abstract}
\end{center}

We present measurements of a transmission-line network, designed
for cloaking applications in the microwave region. The network is
used for channelling microwave energy through an electrically
dense array of metal objects, which is basically impenetrable to
the impinging electromagnetic radiation. With the designed
transmission-line network the waves emitted by a source placed in
an air-filled waveguide, are coupled into the network and guided
through the array of metallic objects. Our goal is to illustrate
the simple manufacturing, assembly, and the general feasibility of
these types of cloaking devices.

\vspace{1.0cm}

\noindent \textit{Keywords:} Transmission-line network; scattering
cross section; electromagnetic cloak.

\vspace{1.0cm}

\section{Introduction}

The interest in different types of devices and materials for
dramatic reduction of the total scattering cross sections of
arbitrary or specific objects, has gained a large amount of
interest after the publication of recent
papers~\cite{Leonhardt,Pendry,Alu,Schurig}. Earlier, the subject
of hiding objects or particles from the surrounding
electromagnetic fields was studied over the recent decades by many
others as well, e.g.,~\cite{Kerker,Chew,Sihvola,Kildal,Greenleaf}.

Recently, we have proposed an alternative approach to cloaking of
objects composed of electrically dense arrays of small inclusions
(in principle, these inclusions can be composed of arbitrary
materials)~\cite{Alitalo_cloak_TAP,Alitalo_cloak_iWAT08}. Since
these objects can be two-dimensional or even three-dimensional
interconnected meshes of e.g. metallic rods, practical
applications of these types of cloaks include hiding strongly
scattering objects such as support structures situated close to
antennas, creating filters (a ``wall'' or a slab letting through
only a part of the spectrum of the incoming field), etc. Also, as
it has been recently proposed, these networks offer a simple way
of creating new types of matched lenses especially for microwave
applications~\cite{Alitalo_TLlens}.

The goal of this paper is to experimentally demonstrate the simple
manufacturing and assembly of the previously proposed
transmission-line structure, where the transmission lines
composing the network are realized as parallel metal strips. By
measurements we confirm the previously predicted matching of the
network with free space, as well as a possibility of transmission
of fields through an electrically dense mesh of metal objects.
These results demonstrate the benefits of this simple approach to
cloaking and present an easy way to measure the performance of
these types of periodic structures.

\section{Transmission-line network}

The transmission-line network that is used here is the same as
designed in Ref.~\cite{Alitalo_cloak_TAP}, with the optimal
impedance matching with free space observed around the frequency
of 5.5~GHz. For this design, the matching with free space and the
cloaking phenomenon were verified with full-wave
simulations~\cite{Alitalo_cloak_TAP}. In this paper we demonstrate
the simple manufacturing and assembly of this type of structure by
choosing to use a two-dimensional periodic transmission-line
network with a square shape and 16~$\times$~16 unit cells in the
network. The edges of the network are connected to a ``transition
layer'' (parallel metal strips gradually enlarging from the ends
of the network), as proposed in Ref.~\cite{Alitalo_cloak_TAP}.

\begin{figure} [t!]
\centering \subfigure[]{\epsfig{file=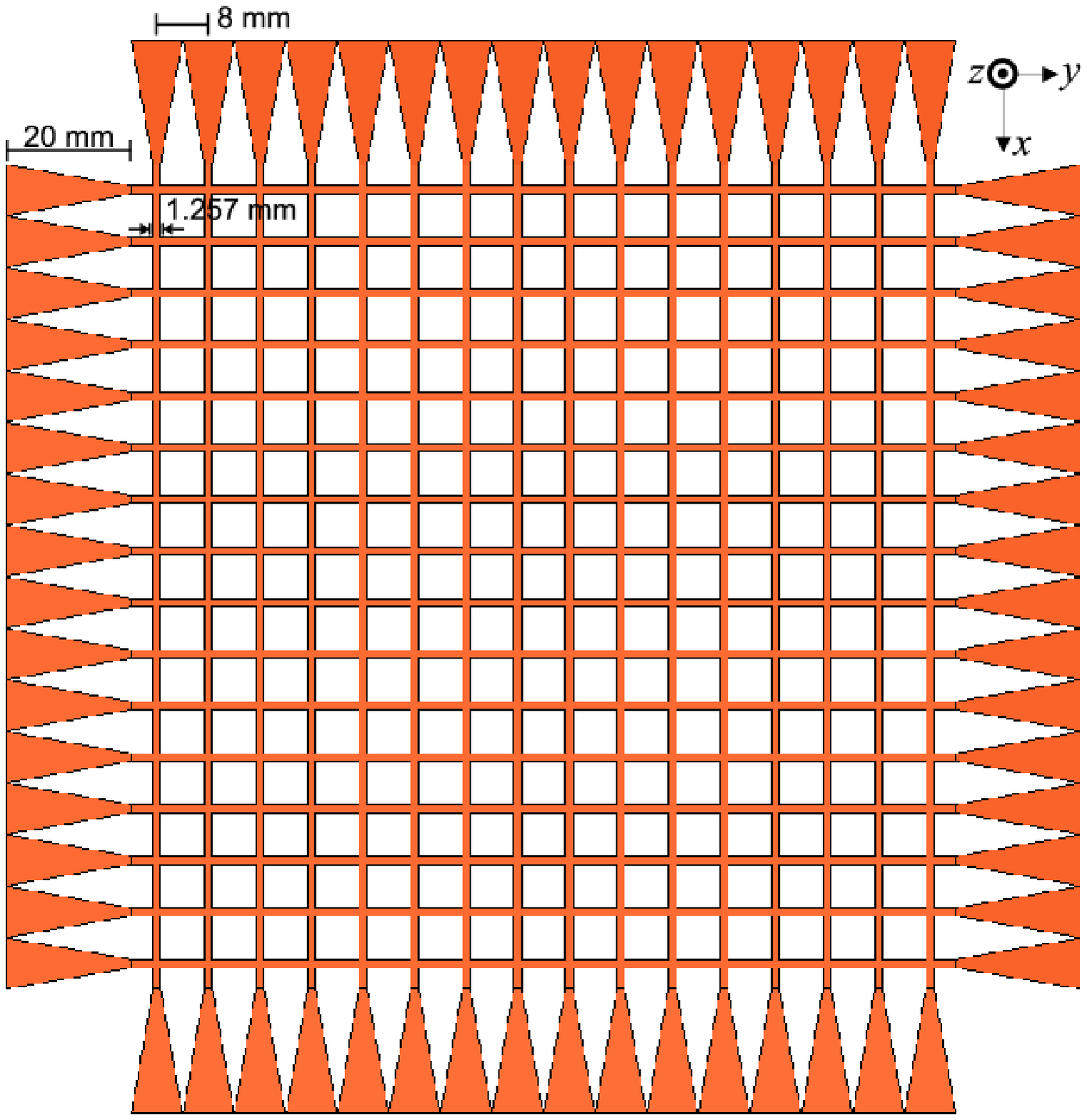,
width=0.49\textwidth}} \subfigure[] {\epsfig{file=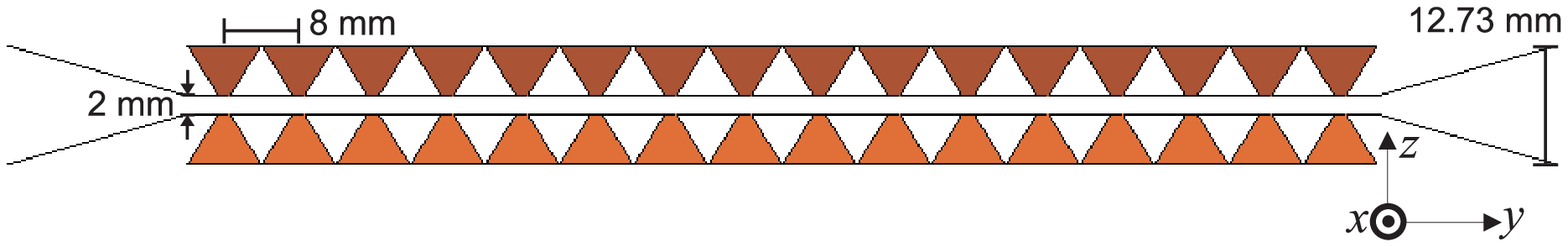,
width=0.49\textwidth}} \caption{Color online. Illustration of the
designed transmission-line network. (a)~Network in the $xy$-plane.
(b)~Network in the $yz$-plane.} \label{network}
\end{figure}

By inserting the designed network inside a metallic parallel-plate
waveguide (with the plates lying in the $xy$-plane), we
effectively realize the same situation as would occur with an
infinitely periodic array of networks with the periodicity along
the $z$-axis, since here the electric field is assumed to be
parallel to the $z$-axis (as in the example case that was studied
previously~\cite{Alitalo_cloak_TAP}). See Fig.~\ref{network} for
an illustration of the transmission-line network with all the
necessary dimensions of the structure included.

The structure shown in Fig.~\ref{network} was manufactured by
etching from a thin copper plate. The network can be simply
assembled from two similar profiles (each profile as shown in
Fig.~\ref{network}a) just by placing them on top of each other, as
shown in Fig.~\ref{network}b. Ideally, the volume between these
two metal objects should be free space~\cite{Alitalo_cloak_TAP}.
Here, due to practical reasons, we have placed small pieces of
styrofoam (with material properties very close to those of free
space) between the metal strips. For assembly purposes, pieces of
styrofoam are placed also on top and below the metal strips for
support. These styrofoam pieces naturally do not affect the
propagation properties of the transmission lines since the fields
are mostly confined between the parallel strips.

In Ref.~\cite{Alitalo_cloak_TAP} the reference object, i.e., the
object that we wanted to cloak (hide) from the surrounding
electromagnetic fields, was an array of infinitely long perfectly
conducting rods (that fit inside the neighboring transmission
lines of the network). Here, we use a similar periodic structure
as a reference object through which we want to guide the fields.
The individual inclusions of this reference object are metal
cylinders (parallel to the $z$-axis) with the same height as the
network ($\sim$13~mm). The diameter of these cylinders is 4~mm and
there are a total of 15~$\times$~15$=225$~cylinders in the array.

\section{Measurement setup}

The measurement setup that is used here is similar to the one
presented in Ref.~\cite{Schurig}. With our measurements we
effectively simulate an infinitely periodic structure with the
periodicity in the vertical ($z$-) direction, by introducing a
measurement cell consisting of a parallel-plate waveguide with its
metallic plates lying in the $xy$-plane. Because of the image
principle (the electric fields are assumed to be mostly parallel
to the $z$-axis inside the waveguide), we can thus measure only
one period of the structure. The difference between the
measurement setup used here and the one in Ref.~\cite{Schurig} is
that here the upper plate of the waveguide is formed by a dense
wire mesh, that lets through a fraction of the field inside the
waveguide, instead of having a solid metallic upper plate with a
hole for the probe, as was used in Ref.~\cite{Schurig}. The mesh
that we use here is the same as the one used in
Ref.~\cite{Maslovski}, i.e., the mesh is a thin copper plate, in
which square holes of size 4~mm~$\times$~4~mm have been etched
with the period of the holes being 5~mm. A small part of the field
gets through this mesh and we can measure that field with a probe
placed on top of the waveguide~\cite{Maslovski}.

To simplify the measurement, we excite a cylindrical wave inside
the waveguide with a feed probe (a coaxial probe placed inside the
waveguide) and measure the transmission from this probe to the
other probe (``measurement probe'', placed on top of the
waveguide) with a vector network analyzer (VNA, Agilent E8363A).
The use of the metal mesh as a part of the top plate of the
waveguide (rather than using a probe inside the waveguide) ensures
that the measurement probe does not disturb the field inside the
waveguide. The measurement probe on top of the waveguide is
stationary, and the waveguide is moved with a PC-controlled
scanner, synchronized with the VNA for precise measurements in the
wanted coordinate positions. These points where the measurement of
the complex $\rm S_{21}$-parameter are taken with the VNA, can be
arbitrarily chosen with the PC-program running the scanner. All
the measurements presented in this paper were done with the steps
of 5~mm. As the measured area is 240~mm~$\times$~100~mm, we will
have the complex $\rm S_{21}$ measured at 49~$\times~21=1029$
different points in the $xy$-plane.

The probe that we use here is a monopole oriented along the
$z$-axis and positioned approximately at 3~mm away from the metal
mesh. The probe is intentionally poorly matched at the frequencies
of interest (5~GHz~--~6~GHz) in order to make sure that the
measurement probe does not disturb the fields inside the
waveguide. The high dynamic range of the VNA makes sure that we
can measure the electric field distribution inside the waveguide
even with this poorly matched probe.

\begin{figure} [t!]
\centering \subfigure[]{\epsfig{file=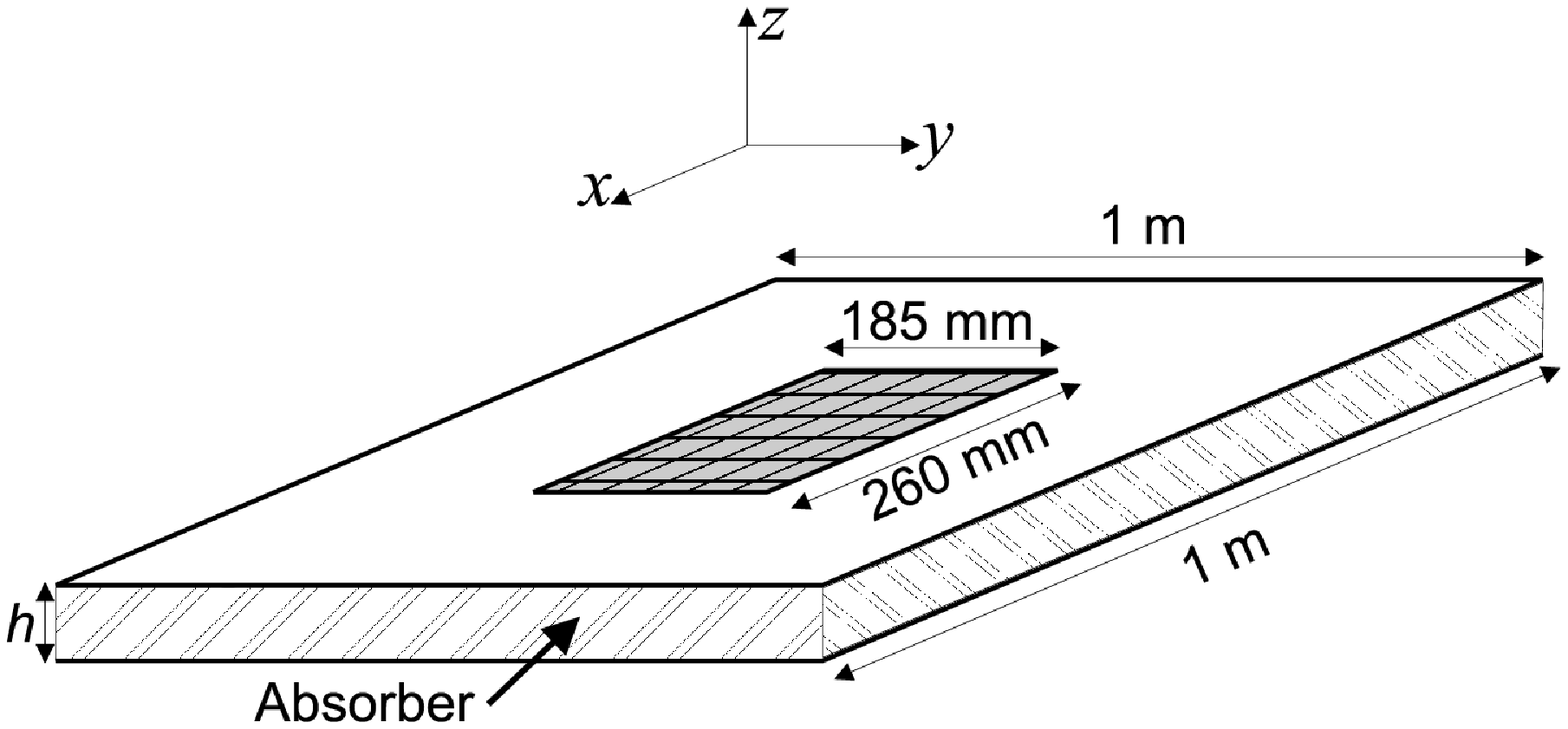,
width=0.49\textwidth}} \subfigure[] {\epsfig{file=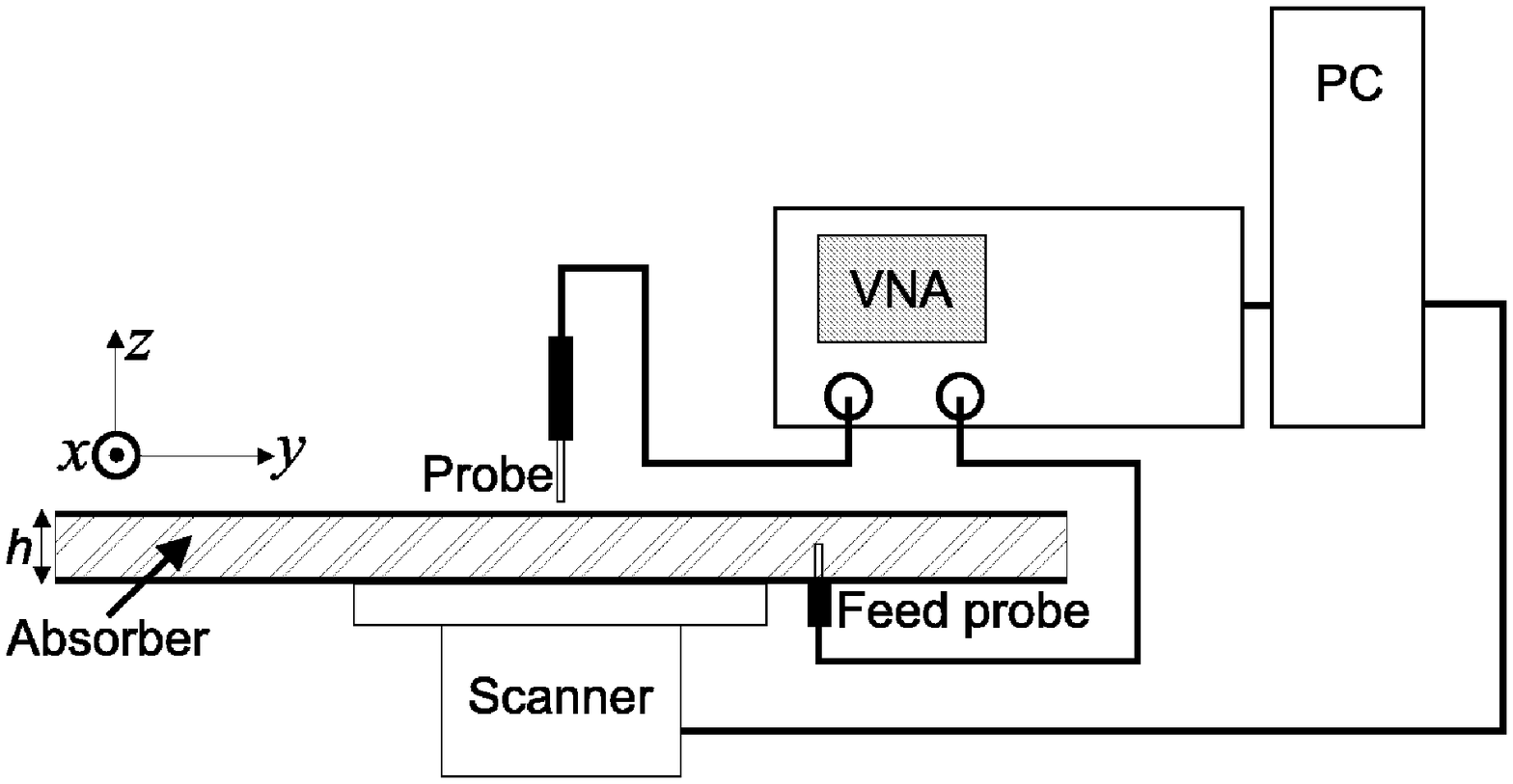,
width=0.49\textwidth}} \caption{Illustration of the measurement
setup. (a)~Waveguide with a metal mesh in the upper plate.
(b)~Measurement system with a VNA connected to the feed and
measurement probes (the measurement probe is stationary) and a PC
controlling the scanner which moves the waveguide in $x$- and
$y$-directions.} \label{setup}
\end{figure}

\begin{figure} [t!]
\centering \subfigure[]{\epsfig{file=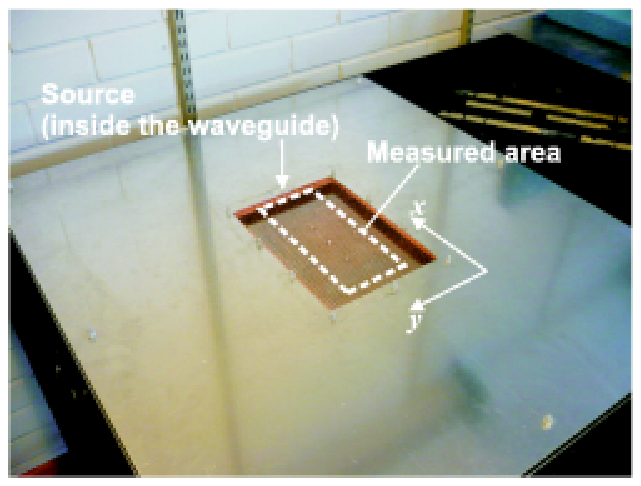,
width=0.45\textwidth}} \subfigure[] {\epsfig{file=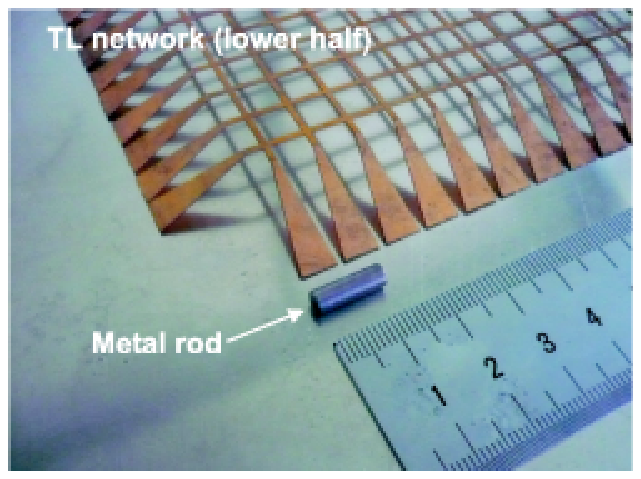,
width=0.45\textwidth}} \caption{Color online. (a)~Photograph of
the measurement setup, showing the aluminium parallel-plate
waveguide and a copper mesh placed in the center of the top plate
of the waveguide. (b)~Photograph of the lower half of the TL
network together with one metallic rod of the reference object.}
\label{photo}
\end{figure}

See Fig.~\ref{setup} for an illustration of the measurement setup.
The parallel-plate waveguide that we use here has the width and
length of 1~m, and the height $h=13$~mm (ideally $h$ should be
equal to 12.73~mm~\cite{Alitalo_cloak_TAP}). A part of the upper
plate is removed from the center for the placing of the metal mesh
(the area above which we want to measure the field distributions).
The area of this mesh is 260~mm~$\times$~185~mm. The measurable
area is further restricted by the used scanners. We have used two
scanners, one with the movement limited to 300~mm ($x$-axis) and
one with the movement limited to 100~mm ($y$-axis). The area to be
measured has been decided to be 240~mm~$\times$~100~mm, centered
in the mesh area. The feed probe is positioned in the center of
the mesh along the $y$-direction and just outside the measured
area in the $x$-direction (to have more space between the measured
area and the feed), i.e., the feed probe coordinates are decided
to be $x=250$~mm, $y=50$~mm, with the origin of this coordinate
system being in one corner of the measured area. See
Fig.~\ref{photo} for a photograph of the measurement setup, taken
from the direction of the positive $z$-axis, showing the empty
waveguide and the metal mesh inserted as a part of the upper
plate. The feed probe position and the measured area are also
illustrated in the figure.

The volume between the waveguide plates, surrounding the metal
mesh, is filled by a microwave absorber. The large size of the
waveguide ensures that the reflections from the waveguide edges
are minimized (the absorber thickness in the $x$- and
$y$-directions is approximately five wavelengths or more at the
frequency of 5~GHz).

\section{Measurement results}

Three different measurements were conducted: 1)~an empty
waveguide, 2)~the reference object (array of 15~$\times$~15 metal
cylinders) inside the waveguide, and 3)~the reference object
\textit{and} the transmission-line network inside the waveguide
(with the inclusions of the reference object placed in the space
between the transmission lines of the network). All the
measurements were conducted in the frequency range from 1~GHz to
10~GHz, with the step of 0.025~GHz.

\begin{figure} [t!]
\centering \subfigure[]{\epsfig{file=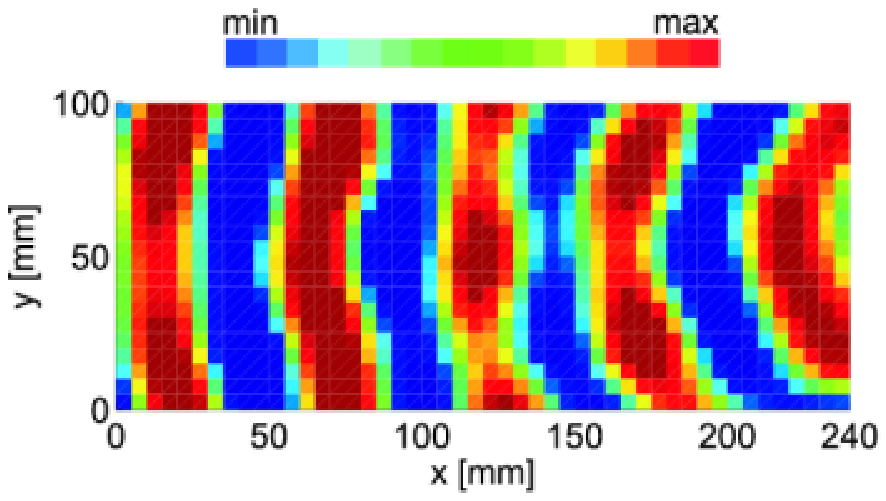,
width=0.45\textwidth}} \subfigure[]
{\epsfig{file=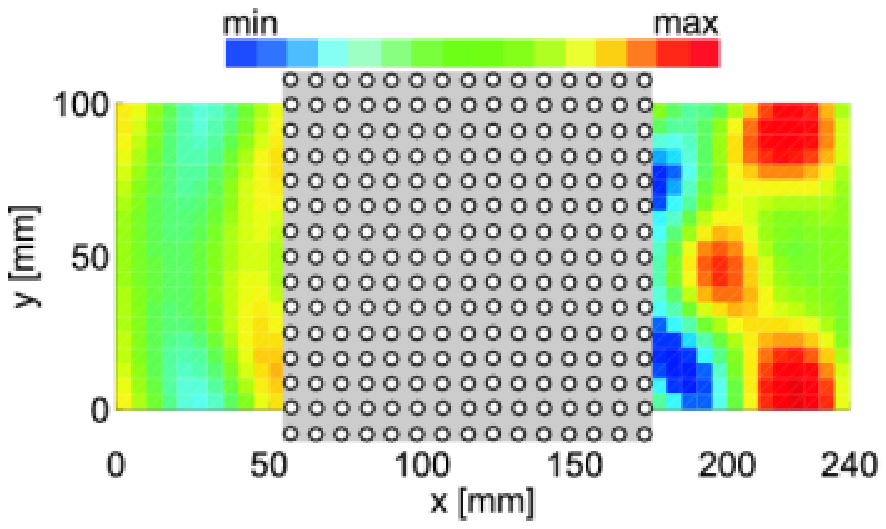, width=0.45\textwidth}}
\subfigure[] {\epsfig{file=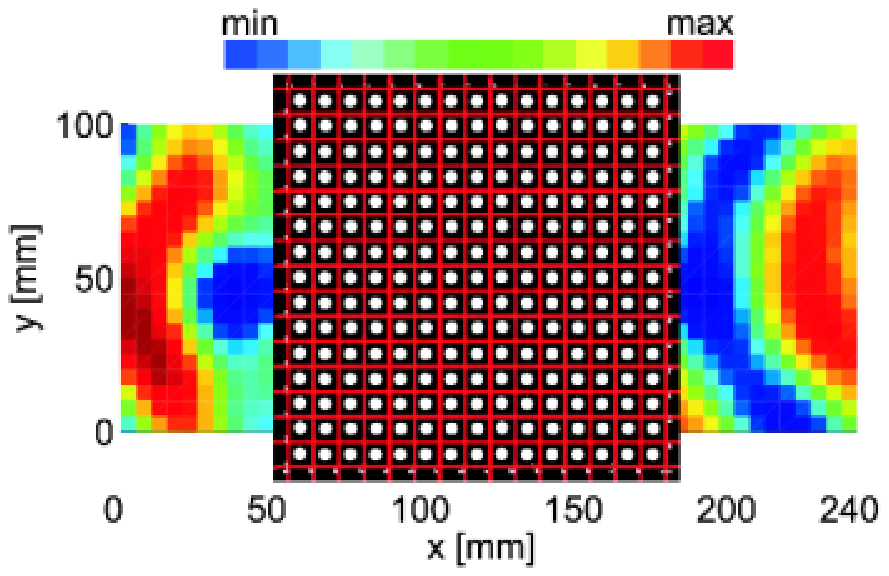,
width=0.45\textwidth}} \caption{Color online. Snapshots of the
measured time-harmonic electric field distributions at 5.85~GHz.
(a) empty waveguide, (b) reference object inside the waveguide,
(c) reference object and the transmission-line network inside the
waveguide. The ``transition layer'' connected to the network is
not shown in (c) for clarity.} \label{THE_5850G}
\end{figure}

In the first case (empty waveguide), the results showed an
expected result: at higher frequencies, i.e., at 5~GHz and up, the
waveform inside the waveguide is close to the waveform produced by
a line source. At lower frequencies, where the waveguide is
electrically smaller, the reflections from the edges start to
affect the field distributions, making them more complicated. See
Fig.~\ref{THE_5850G}a for a snapshot of the measured time-harmonic
electric field distribution at the frequency 5.85~GHz. Some
reflections naturally still occur (mainly from the absorbers), but
it is clear that the wave inside the waveguide resembles a
cylindrical wave emanating from the point $x=250$~mm, $y=50$~mm.

In the second case (reference object inside the waveguide), the
results were again as expected: at the higher frequencies, where
it makes sense to compare the field distributions, the wavefronts
emanating from the source are strongly reflected at the front
boundary of the reference object. The field is seen to ``split''
in the center and move up or down along the reference object side.
See Fig.~\ref{THE_5850G}b for a snapshot of the measured
time-harmonic electric field distribution at the frequency
5.85~GHz.

In the third case (the reference object and the transmission-line
network inside the waveguide) the field pattern on the source side
is seen to be well preserved (as compared to
Fig.~\ref{THE_5850G}a) in a certain frequency band around
5.85~GHz. Also, at the backside of the network, some field is
propagating. At the position of the network (which encompasses the
reference object), no field is measured, since the fields are
confined inside the transmission lines. See Fig.~\ref{THE_5850G}c
for a snapshot of the measured time-harmonic electric field
distribution at the frequency 5.85~GHz.

\begin{figure} [t!]
\centering \subfigure[]{\epsfig{file=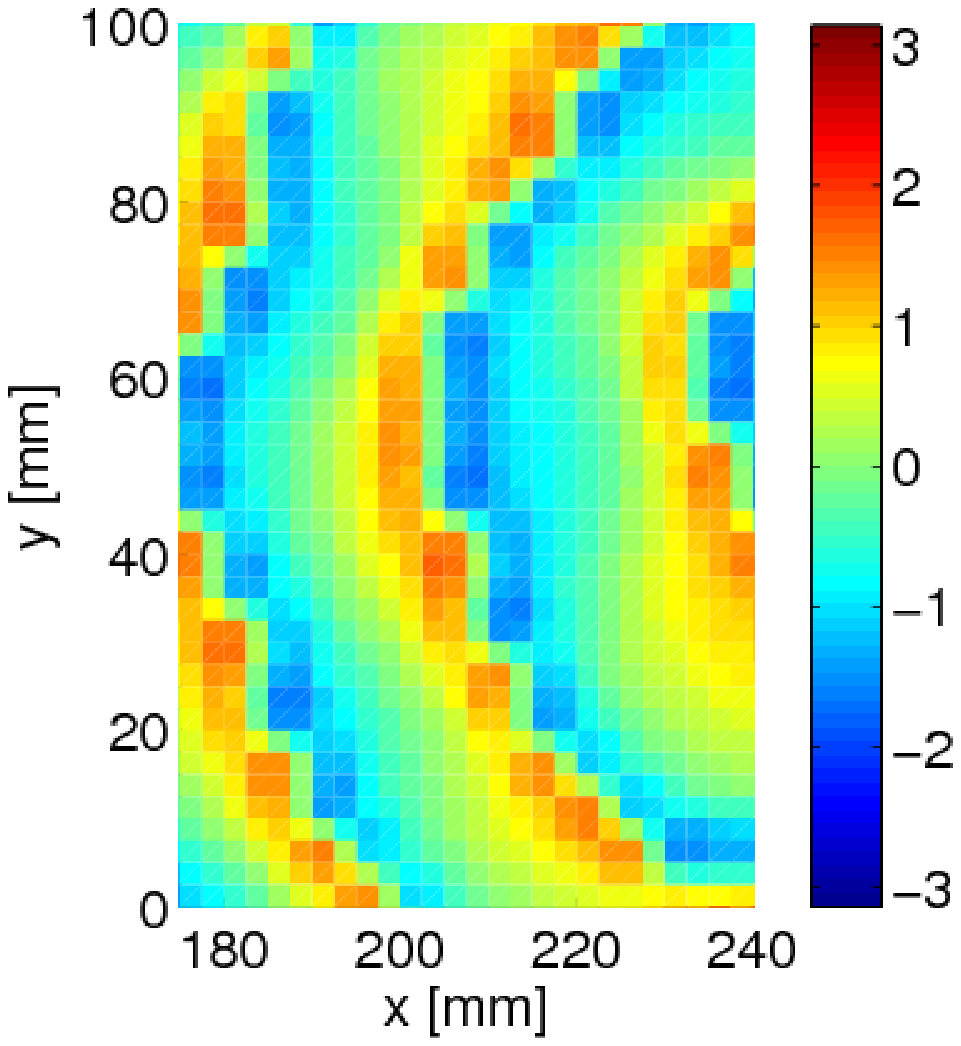,
width=0.25\textwidth}} \subfigure[]
{\epsfig{file=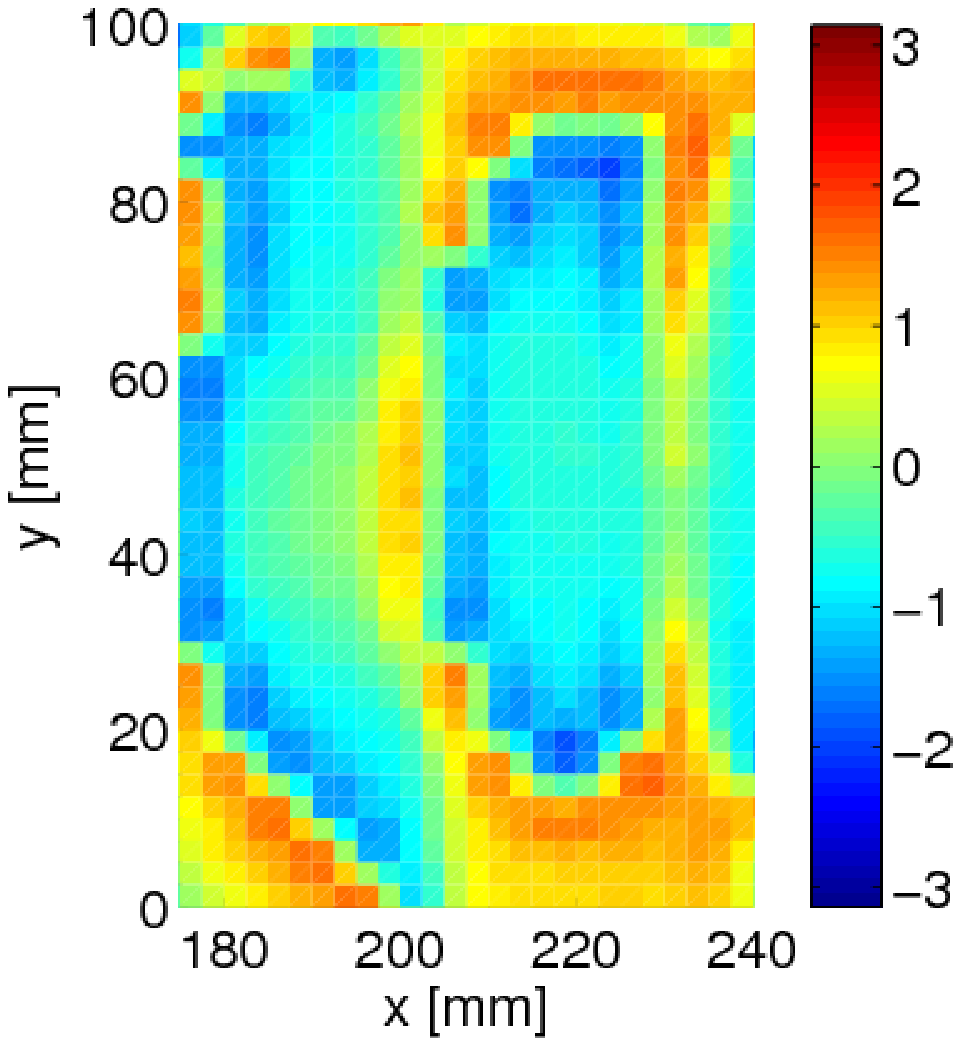, width=0.25\textwidth}}
\subfigure[] {\epsfig{file=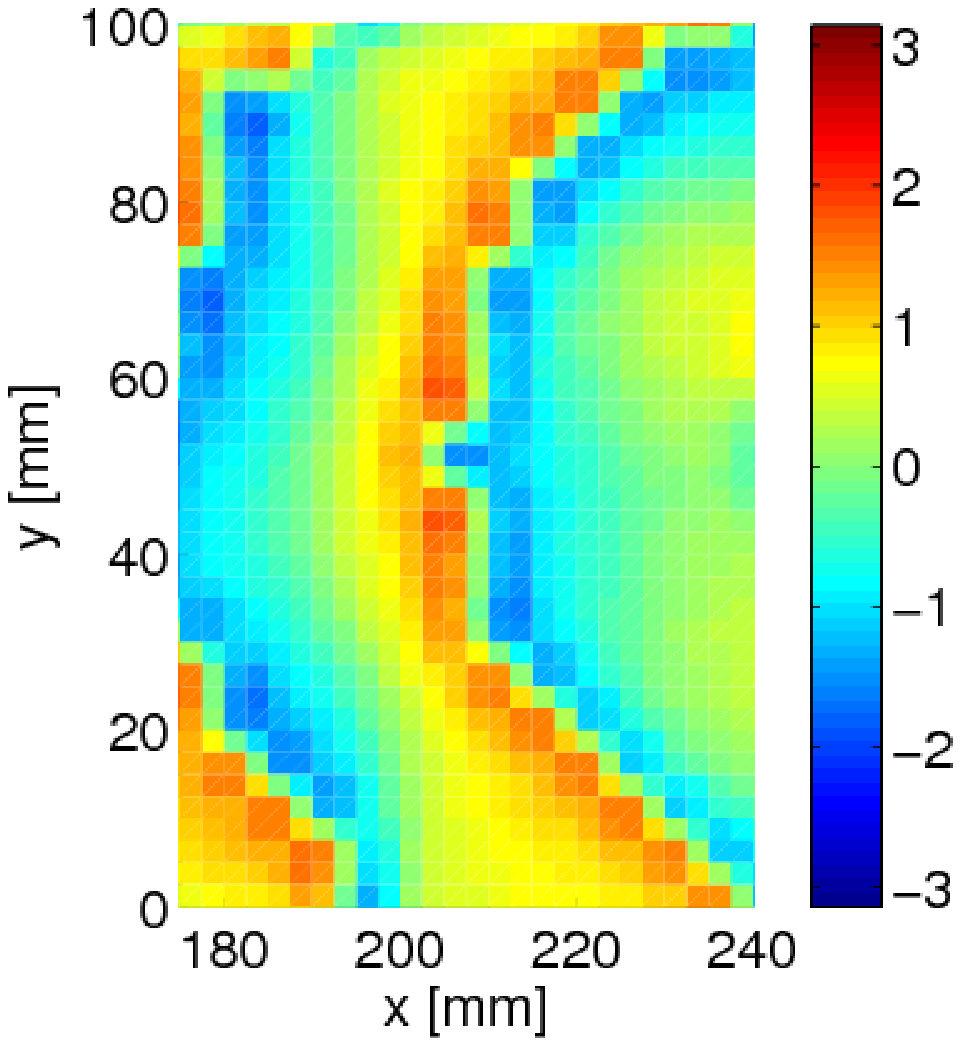,
width=0.25\textwidth}} \caption{Color online. Phase of the
measured electric field distribution at 5.85~GHz. (a) empty
waveguide, (b) reference object inside the waveguide, (c)
reference object and the transmission-line network inside the
waveguide. The phase distributions in each figure are interpolated
from the corresponding measurement data for clarity.}
\label{phase_5850G}
\end{figure}

To further study the differences between the situations with and
without the network, the phase distributions, calculated from the
measured complex field data, are shown in Fig.~\ref{phase_5850G},
plotted only in the area between the reference object/network and
the feed probe, i.e., in the area $x=175$~mm~...~$x=240$~mm. As
compared to the empty waveguide (Fig.~\ref{phase_5850G}a), the
case with the bare reference object, Fig.~\ref{phase_5850G}b,
looks very different. This is due to the strong reflections from
the front edge of the reference object. When the transmission-line
network is placed inside the waveguide, together with the
reference object, we see that the resulting phase distribution
again is close to the one in Fig.~\ref{phase_5850G}a.

\begin{figure} [t!]
\centering \subfigure[]{\epsfig{file=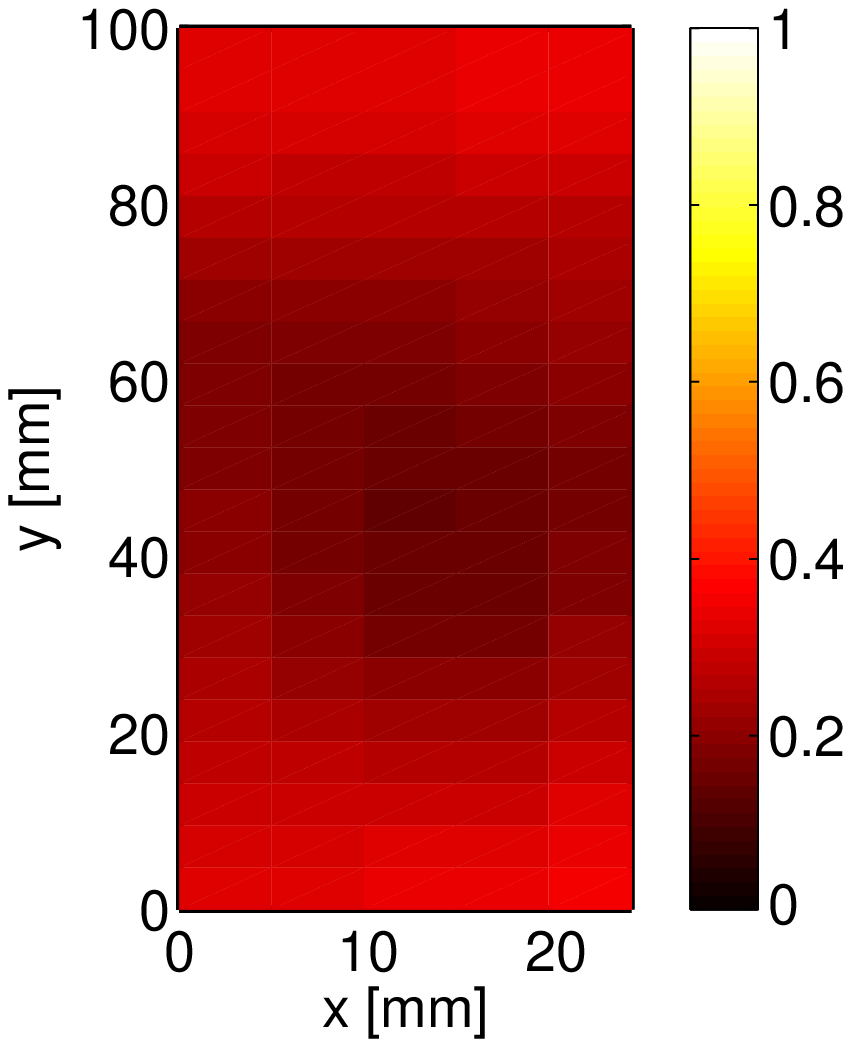,
width=0.23\textwidth}} \subfigure[]
{\epsfig{file=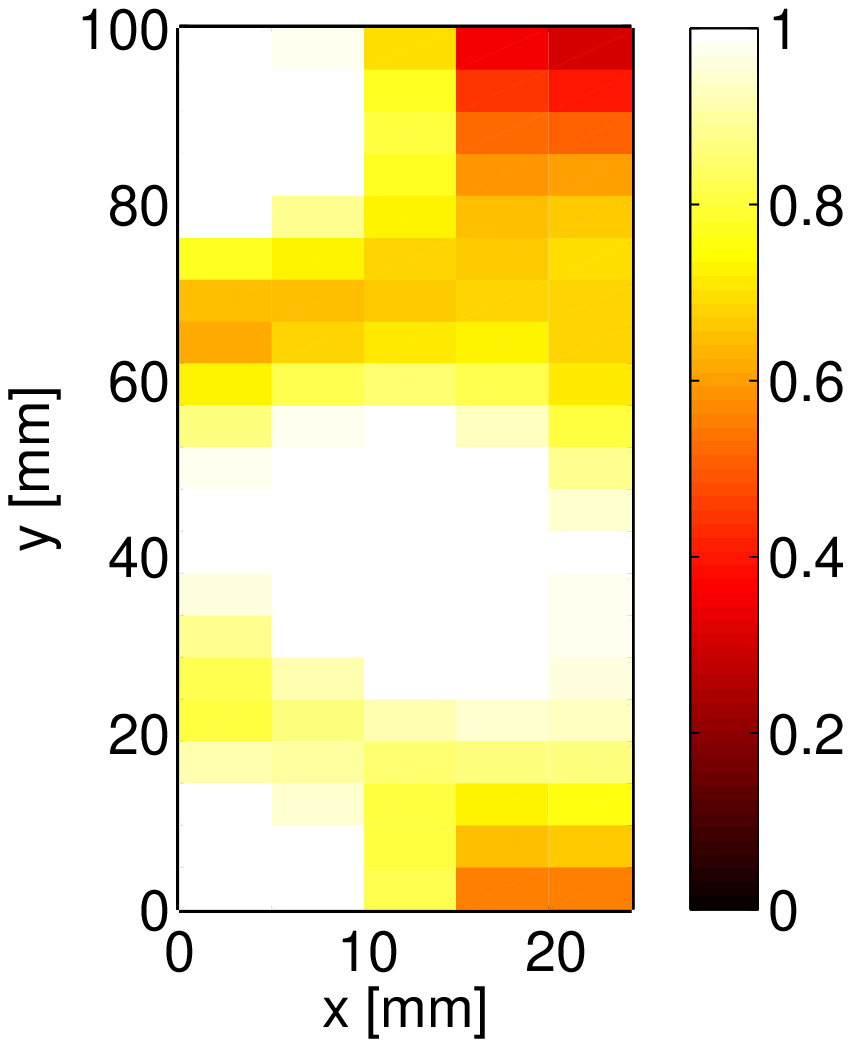,
width=0.23\textwidth}} \caption{Color online. Absolute value of
the measured electric field distribution at 5.85~GHz, normalized
to the maximum field value measured in front of the reference
object/network (in the region $x=175$~mm~...~$x=240$~mm).
(a)~reference object inside the waveguide, (b)~reference object
and the transmission-line network inside the waveguide.}
\label{absE_5850G}
\end{figure}

A more illustrative measure for the operation of the network is to
compare the absolute value of the field \textit{behind} the
reference object with and without the network in place (i.e, in
the area $x=0$~mm~...~$x=50$~mm). These results are shown in
Fig.~\ref{absE_5850G}a and Fig.~\ref{absE_5850G}b, for the case
without the network and with the network, respectively. As
demonstrated by Fig.~\ref{absE_5850G}, the field amplitude behind
the reference object is strongly suppressed (as compared to the
field amplitude in front of this object). With the
transmission-line network in place, the field amplitude behind the
reference object/network is practically the same as in front.

The size of the network in this example case is comparable to the
wavelength, so the inherent difference of the phase velocities
between the wave propagating inside the network and the wave in
free space, results in a change, or distortion of the impinging
cylindrical waveform, as can be seen from Fig.~\ref{absE_5850G}b.
The reasons for this difference in the phase velocity are
discussed in Ref.~\cite{Alitalo_cloak_TAP}. Also, as the sharp
edges of the square-shaped network are close to the measured area,
scattering from these edges distorts the waveform. Note that the
previously simulated ``cloak slab''~\cite{Alitalo_cloak_TAP} was
much wider than the network measured here.

To obtain efficient cloaking, i.e., to preserve the waveform of
the incident wave both in front and behind the object (to reduce
the total scattering cross section), one clearly needs to use a
cloak which is electrically small enough and which does not have
strong irregularities in its shape, not to cause significant
forward scattering. It is also possible to use an electrically
large cloak, dimensions of which are properly designed for a
specific frequency range so that the desired reduction of the
forward scattering is achieved~\cite{Alitalo_cloak_TAP}. When the
incidence angle of the impinging radiation is not known, the cloak
naturally needs to be symmetric, i.e., cylindrical in the
two-dimensional case or spherical in the three-dimensional
case~\cite{Alitalo_cloak_TAP,Alitalo_cloak_iWAT08}. Also, what is
important in the case when the angle of the incident radiation is
not known, the transmission-line network needs to be isotropic in
the frequency range where most efficient cloaking is needed, i.e.,
the network period must be small enough as compared to the
wavelength, as e.g. in the cylindrical cloak studied in
Ref.~\cite{Alitalo_cloak_iWAT08}.


\section{Conclusions}

We have presented a prototype of a previously proposed and
designed transmission-line network for cloaking purposes. In this
paper we have demonstrated the benefits of this approach, such as
the simple manufacturing, and confirm by measurements the
predicted impedance matching with free space and the resulting
field propagation through such a network (with a periodic array of
metallic rods placed between the transmission lines of the
network). The results are compared with a measurement of an empty
measurement cell and with a measurement of the periodic array of
metallic rods, effectively behaving as an impenetrable wall at the
frequencies of interest.

\section*{Acknowledgements}

This work has been partially funded by the Academy of Finland and
TEKES through the Center-of-Excellence program and partially by
the European Space Agency (ESA--ESTEC) contract no. 21261/07/NL/CB
(Ariadna program). The authors wish to thank Mr.~E.~Kahra and
Mr.~L.~Laakso for valuable help with manufacturing of the
prototype and with building of the measurement setup. P.~Alitalo
acknowledges financial support by the Finnish Graduate School in
Electronics, Telecommunications, and Automation (GETA), Tekniikan
Edist\"amiss\"a\"ati\"o (TES), and the Nokia Foundation.


\end{document}